\documentclass[aps,showpacs,floatfix,superscriptaddress,twocolumn]{revtex4-1}
\usepackage{graphicx}
\usepackage{amssymb}
\usepackage{graphics, color}
\usepackage{amsmath}
\usepackage{verbatim}
\usepackage{calc} 
\usepackage{esint}
\usepackage{hyperref}

\usepackage[T1]{fontenc}
\usepackage[latin9]{inputenc}
\usepackage{float}

\def\gsim{\, \rlap{$>$}{\lower 1.1ex\hbox{$\sim$}}\,}
\def\lsim{\, \rlap{$<$}{\lower 1.1ex\hbox{$\sim$}}\,}

\definecolor{orange}{rgb}{1,0.5,0}

\newcommand{\be}{\begin{equation}}
\newcommand{\ee}{\end{equation}}
 \newcommand{\bal}{\begin{align}}
 \newcommand{\eal}{\end{align}}
\newcommand{\ben}{\begin{equation*}}
\newcommand{\een}{\end{equation*}}
\newcommand{\bea}{\begin{eqnarray}}
\newcommand{\eea}{\end{eqnarray}}
\newcommand{\bean}{\begin{eqnarray*}}
\newcommand{\eean}{\end{eqnarray*}}
\newcommand{\bes}{\begin{subequations}}
\newcommand{\ees}{\end{subequations}}

\def\beq{\begin{equation}}
\def\eeq{\end{equation}}

\begin{document}
\title{Dissipation in a topological Josephson junction}
\author{Paul Matthews}
\affiliation{CIC nanoGUNE, Tolosa Hiribidea 76, 20018 Donostia-San Sebastian, Spain}
\affiliation{University of Cambridge, JJ Thomson Avenue, Cambridge, CB3 0HE, UK}
\author{Pedro Ribeiro}
\affiliation{Max Planck Institute for the Physics of Complex Systems - N\"othnitzer Str. 38, D-01187 Dresden, Germany}
\affiliation{Max Planck Institute for Chemical Physics of Solids - N\"othnitzer Str. 40, D-01187 Dresden, Germany}
\affiliation{CFIF, Instituto Superior T{\'e}cnico,
Universidade T{\'e}cnica de Lisboa, Av. Rovisco Pais, 1049-001 Lisboa, Portugal}
%\affiliation{CFIF, Instituto Superior T\'ecnico, TU Lisbon, Av. Rovisco Pais, 1049-001 Lisboa, Portugal}
\author{Antonio M. Garc\'{\i}a-Garc\'{\i}a}
\affiliation{TCM Group, Cavendish Laboratory, University of Cambridge, JJ Thomson Avenue, Cambridge, CB3 0HE, UK}
\affiliation{CFIF, Instituto Superior T{\'e}cnico,
Universidade T{\'e}cnica de Lisboa, Av. Rovisco Pais, 1049-001 Lisboa, Portugal}
%\author{Masaki Tezuka}
%\affiliation{Department of Phys, Kyoto University, Kitashirakawa, Sakyo-ku, Kyoto 606-8502, Japan}
\begin{abstract}
Topological features of low dimensional superconductors have caused a lot of excitement recently because of their broad range of applications in quantum information and their potential to reveal novel phases of quantum matter.
A potential problem for practical applications is the presence of phase-slips that break phase coherence. Dissipation in non-topological superconductors suppresses phase-slips and can restore long-range order. Here we investigate the role of dissipation in a topological Josephson junction. We show that the combined effects of topology and dissipation keeps phase and anti-phase slips strongly correlated so that the device is superconducting even under conditions where a non-topological device would be resistive. The resistive transition occurs at a critical value of the dissipation which is four times smaller than that expected for a conventional Josephson junction. We propose that this difference could be employed as a robust experimental signature of topological superconductivity.
\end{abstract}
\date{\today}
\maketitle
\section{Introduction}
The Josephson effect \cite{14,1} not only reveals the central role played by the phase of the superconducting 
order parameter, but it is also key in many applications of superconductivity in electronics, such as 
superconducting quantum interference devices (SQUID) that can measure exceedingly small spatial variations of 
a magnetic field.

For bulk samples it can be simply stated as the existence of a current $I = I_c\sin(\phi)$ between two 
superconductors separated by a thin metal or insulator \cite{14,1}, where $\phi \equiv \phi_1-\phi_2$ is the 
phase between the two superconductors and \cite{bara} $I_c \approx \frac{\pi \Delta} {2R_Ne}$ is the so-
called critical current with $e$ the electron charge and $\Delta$ the zero temperature superconducting gap. 
The normal state resistance is given by $R_N =\hbar/ \left[ 4\pi e^2|t|^2N_L(0)N_R(0) \right]$ 
with  $t$ the tunnelling matrix element, and $N_{L/R}(0)$ the normal state electronic density of states at the Fermi energy of the left/right superconductor.

As the system size decreases charging effects induce fluctuations in the phase which can potentially destroy phase coherence. At the same time there are different mechanisms of dissipation \cite{caldeira} which can quench these fluctuations and restore long range order.

Ambegaokar et al. \cite{2} derived an action from a microscopic Hamiltonian \cite{2} that includes the Josephson coupling ($\propto I_c$), the intrinsic quasiparticle tunnelling between the two identical 
 superconductors, and a finite capacitance due to the Coulomb interaction across the barrier. The (Euclidean) final action of \cite{2} takes the form,
\begin{eqnarray}
\label{actschoen}
S[\phi] = \int_0^{\beta\hbar}\frac{d\tau}{\hbar}\left[\frac{C}{2}\left(\frac{\hbar}{2e}\partial_\tau \phi\right)^2-\frac{I_c \hbar}{2e}\cos(\phi)\right] \nonumber \\ +2\int_0^{\beta \hbar}d\tau d\tau' \alpha(\tau-\tau')\times \sin^2\left[\frac{\phi(\tau)-\phi(\tau')}{4}\right] 
\end{eqnarray}
with $C$ the mutual capacitance between the two superconductors. The last term of Eq. (\ref{actschoen}) describes dissipation across the junction. In the case of quasiparticle induced dissipation the kernel $\alpha(\tau) \approx \frac{\hbar}{2\pi e^2R_N}\frac{1}{\tau^2}$ decays as a power-law for short times $\tau \ll \hbar/\Delta$ and exponentially $\alpha(\tau) \propto e^{-2\Delta \tau/\hbar}$ for $\tau \gg \hbar/\Delta$. Therefore for long times $\tau \gg \hbar/\Delta$ tunnelling of quasiparticles plays a relatively minor role in the phase dynamic which may be seen as a simple renormalization of the capacitance $C$ \cite{2}.

At zero temperature, and for short times $\tau \ll \hbar/\Delta$, the dissipative term of Eq.(\ref{actschoen}) resembles the one introduced by Caldeira and Leggett \cite{caldeira} to describe Ohmic dissipation in a quantum system induced by a linear coupling to a bath of harmonic oscillators, 
\begin{eqnarray}
\label{caldeira1}
S_{diss}[\phi] =\frac{\eta}{4\pi}\int d\tau d\tau' \left[\frac{\phi(\tau)-\phi(\tau')}{\tau-\tau'}\right]^2
\end{eqnarray}
where,  in the classical limit,  this source of dissipation corresponds to a Langevin equation with $\eta$ the friction coefficient. 
This action also describes an Ohmic resistance across a Josephson junction \cite{22}. 
Indeed, replacing the sine term by its argument, the last term in 
Eq.(\ref{actschoen}) is equivalent to Eq.(\ref{caldeira1}) providing that $\eta =\frac{\hbar}{4e^2R_N} \propto R_q/R_N$
where $R_q=h/e^2$ is the quantum resistance. 
Physically the replacement of the sine by its argument is a valid approximation only in the limit in which capacitance effects are not very strong so that the charge can still be considered a continuous classical variable \cite{18}. Ohmic dissipation can also be induced \cite{gil1} by the proximity to normal metals or to normal-state conducting channels. 

The action in Eq.(\ref{actschoen}) contains a potential term with an infinite set of degenerate minima in 
addition to the kinetic and dissipative contributions. 
Tunnelling among different minima lowers the ground state energy and therefore plays an important role in the 
description of the system. 
In this context a tunnelling event of the phase between two consecutive minima, also referred to as a phase-slip 
or instanton, shifts the phase of the order parameter by a multiple of $2\pi$. 
These large quantum fluctuations have the potential to break phase coherence in the system. 
At the same time it is well known that Ohmic dissipation suppresses tunnelling \cite{caldeira}.

The interplay between these two mechanisms has been thoroughly investigated in the literature both for a 
double well \cite{19,20} and for a periodic potential (sine-Gordon) \cite{4,21,22}. 
In the semi-classical limit, instanton solutions to the non-dissipative action ($\eta = 0$) are a good 
approximation of the total action solution.

In the limit where instantons are dilute, the partition function can be calculated by integrating over all 
multi-instanton paths. 
By performing a scaling analysis of the resulting expression, renormalization group (RG) equations are 
derived that ultimately provide a good qualitative picture of the phase diagram \cite{23,25}. 
For a periodic monochromatic potential this was done exploiting the mapping into a 
one dimensional Ising model with inverse square interactions \cite{26,27} also known to be equivalent to the two dimensional log-gas \cite{28} and to the two dimensional XY model \cite{29}.

The mapping into these  models, for which the phase diagram is well known, confirmed that at zero 
temperature, there is a continuous phase transition for a finite value of $\eta=\eta_c$ from a phase where phase 
slips destroy global superconductivity, to a phase of strong dissipation where tunnelling is suppressed and 
the phase of the order parameter stays in a single potential minimum. 
Technically the dissipative term introduces instanton-(anti-)instanton correlations which eventually fully suppress tunnelling of the phase for $\eta \geq \eta_c$.

A transition only occurs for dissipation with a sufficiently slow 
power-law decay kernel. As mentioned previously this is not the case for intrinsic quasiparticle dissipation 
\cite{2} whose kernel  $\alpha(\tau) \propto \exp(-2 \tau\Delta/\hbar)$ decays exponentially for long times 
so that the (anti-)instanton interaction is short-range and therefore is not enough to stabilize global 
superconductivity. In that case the effect of dissipation is simply to weaken charging effects by 
renormalizing the capacitance. Phase slips will likely still create a local voltage fluctuation making the 
junction resistive. The ultimate reason for this behaviour can be traced back to the energy gap $2\Delta$ 
that severely penalizes quasiparticle tunnelling.

The recent claimed observation \cite{mourik} of Majorana fermions in InSb nanowires and its potential 
relevance in the context of quantum information \cite{4,37,kwon,48} has boosted research in topological 
superconductivity. Especially for applications it is of interest to explore dissipative effects in 
materials \cite{4,32} characterized by zero-energy sub-gap excitations. The existence of superconductors with 
topological features was first speculated in $\nu =5/2$ fractional quantum hall states \cite{33} and then on 
the edges of effectively spinless systems with triplet pairing symmetry \cite{34,1}. Later \cite{35} it was 
proposed to realise topological superconductivity with surface states using the proximity effect between a 
strong topological insulator and an ordinary s-wave superconductor. Further work \cite{36,37} has revealed 
that this requirement can be realised in one-dimensional semiconductor wires. Several other proposals have been put forward recently in order to observe experimentally topological superconductivity \cite{38,39,40,41}.

Here we investigate the role of dissipation in a Josephson junction (JJ) composed of two topological 
superconductors separated by a weak link. Starting from a microscopic Hamiltonian we show that
dissipation in a topological JJ suppresses phase-slips more strongly than in a conventional JJ. We have 
identified a critical value of the dissipation strength, which is four times smaller than in conventional JJ's, above which phase slips are suppressed and a supercurrent is stable.

The paper is organized as follows: In the next section we introduce the model and construct the classical 
instanton solutions to the quantum mechanical action as derived in \cite{3}. We then evaluate the partition 
function to leading order by summing over all instanton contributions in a saddle point analysis. An RG 
approach, similar to the one introduced by Bulgadaev \cite{4} for a JJ with Ohmic dissipation, is employed to 
determine the phase diagram of the topological superconducting device. Results are discussed in section III. 
Finally we draw conclusions in section IV. 
\section{The model}
The physical setup we consider corresponds to a one-dimensional wire where superconductivity is induced by proximity effect and topological features are a consequence of a strong spin-orbit coupling together with a perpendicularly applied magnetic field \cite{45}. The proximity to the nearby bulk superconductor induces an effective attractive density-density interaction between electrons on neighbouring atomic sites. The Josephson junction is modelled by a weak link between the left and right part of the wire. 
For concreteness we assume a simple tight-binding model for the wires in the normal state. 
A simplified Hamiltonian for the system is given by, 
\begin{align}
H= &
  \sum_{n=0,l=R,L} t \left( c_{n,l}^{\dagger} c_{n+1,l} + \text{h.c.} \right) +s \left(c_{0,L}^{\dagger}c_{0,R} + \text{h.c.} \right)
 \nonumber \\&-g\sum_{n,l}c_{n+1,l}^{\dagger}c_{n+1,l}c_{n,l}^{\dagger}c_{n,l}
 \label{eq:H}
\end{align}
where $t$ is the intra-wire hopping, $s$ is the weak link tunnelling and $g$ is the effective coupling constant. 
At the mean-field level with  $\Delta_{n,n+1;l}  = - g \left\langle   c_{n,l} c_{n+1,l}   \right\rangle$, this Hamiltonian recovers a generalized  Kitaev model \cite{48} by the substitution $- g c_{n+1,l}^{\dagger}c_{n+1,l}c_{n,l}^{\dagger}c_{n,l} \to  c_{n+1,l} ^{\dagger}  c_{n,l}^{\dagger} \Delta_{n,n+1;l} +  \bar \Delta_{n,n+1;l} c_{n,l} c_{n+1,l}    +  g^{-1} \bar \Delta_{n,n+1;l}    \Delta_{n,n+1;l}  $. 

As in the non-topological case, the effective low energy theory of the model involves only the difference between the superconducting phases across the weak link  $ \phi = \arg(\Delta_{0,1;L}) -  \arg(\Delta_{0,1;R})$. 
The microscopic derivation of the effective action for the junction follows the Eckern-Schoen-Ambegaokar calculation \cite{2} for a conventional (non-topological) superconductor with an important difference: the presence of a bound-state at the weak link. In the topological case the single particle Green's function can be decomposed into a bound-state and a continuum part. The former represents the effect of the gapped quasiparticles and, as in the non-topological case, can be treated in second order perturbation theory in the weak link hopping magnitude $s$. This contribution yields an effective capacitive term, proportional to $(\partial_\tau \phi)^2$, and the Josephson term, proportional to $\cos(\phi)$ \cite{2}. The bound-state contribution cannot be treated perturbatively and  requires the knowledge of the bound-state wave function. As the bound-state wave function cannot decay to the quasiparticle continuum the occupation of the mixed particle-hole wave function - corresponding to two Majorana modes - is not a dynamic variable, being either empty or occupied.  This problem has been considered by Pekker et al. \cite{3} for the case where the magnitude of the order parameter equals the intra-wire hopping $\left| \Delta \right| = t $, corresponding to a particularly simple form of the bound-state wave function. The appearance of a new $\cos[\phi(\tau)/2]$ term, particularly transparent in the treatment of Ref.\cite{3}, is expected to occur for all values of the intra-wire hopping.  
At zero-temperature, after integration over the fermionic degrees of freedom, the effective Euclidean action is given by,
\begin{align}
S_{0} &=& \int  \left[\frac{ \left(\partial_\tau {\phi}\right)^2}{16E_c}-E_J(1-\cos\phi)\pm \frac{E_M}{2}\cos(\phi/2)\right] d\tau'
\end{align}
which corresponds to the so-called double sine-Gordon action \cite{3} where $E_c$ is the charging energy due to the capacitance which will eventually be renormalised by quasiparticle tunnelling. 
$E_J$ is the Josephson coupling and $E_M$ is the energy associated with the two Majorana fermions localised at the weak link which is proportional to the hopping amplitude $s$ for an electron to tunnel across the junction. 

\begin{figure}
\includegraphics[width=0.99\columnwidth,clip,angle=0]{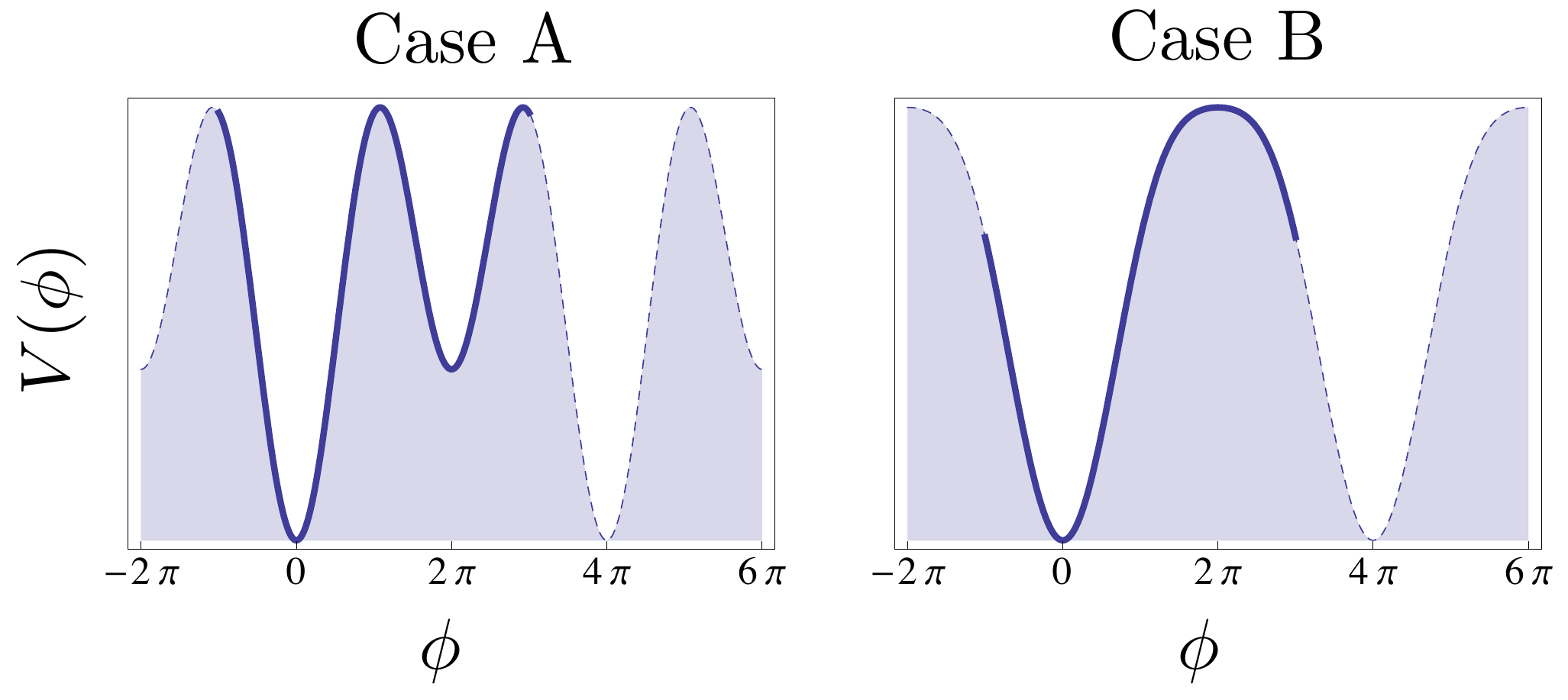}
\caption{Effective potential Eq. (\ref{efe}) for odd parity controlling the phase dynamic of a topological superconducting junction. Case A: $0 < \lambda < 4\mu$ and both a local and a global minimum exist. 
Case B: $\lambda  > 4\mu$ and only a global minimum exists \cite{49}.
} 
\label{fig3}
\end{figure}

The positive (even) and negative (odd) energy states in this setup correspond to whether the bound-state made of the two single Majorana fermions is occupied or empty. 
Here, parity corresponds to the eigenvalue of the number operator of the bound-state \cite{1}. 
This symmetry labels the two lowest energy states of the system. 
Note that, see Fig. \ref{fig3}, the different parities are related by a translation of the potential by $2\pi$ along the $\phi$ axis. 
In the following, without loss of generality, we only treat odd parity and infer the even parity results from the translational symmetry.
Defining $\mu = 8E_CE_J$, $\lambda = 4E_CE_M$ the double sine-Gordon potential (with $\lambda>0$) 
\begin{eqnarray}
V(\phi) &=& \mu[1-\cos(\phi)]+\lambda[1-\cos(\phi/2)]
\label{efe}
\end{eqnarray}
is shown schematically in Fig. \ref{fig3} for two qualitatively different cases  characterized by the existence or not of a local minimum.
%%%%%%%%%%%%%%%%%%%%%%%%%%%%%%%%%%%%%%
%%%%%%%%%%%%%%%%%%%%%%%%%%%%%%%%%%%%%%

We now consider the role of a dissipative term in the topological junction. 
The total action is thus given by, 
\begin{eqnarray}
  S_{top}[\phi] &=& \frac{1}{8E_C} ( S'_0[\phi]+S'_{diss}[\phi] )
\label{top2}
\end{eqnarray}
where 
\begin{eqnarray}
S'_{0} &=& \int  \left\{ \frac{ \left(\partial_\tau {\phi}\right)^2}{2} - V[\phi] \right\} d\tau \nonumber
\end{eqnarray}
and 
$S'_{diss}$ acquires the Caldeira and Leggett \cite{caldeira} form,
\begin{eqnarray}
S'_{diss} &=& \tilde\eta \int\frac{[\phi(\tau)-\phi(\tau')]^2}{(\tau-\tau')^2}d \tau d\tau' \nonumber
\end{eqnarray}
where $\tilde \eta =8\eta E_c$. Note that for quasiparticle dissipation $\eta = \frac{\hbar}{16\pi e^2 R_N}$ while it is a free parameter for a generic resistive Ohmic shunt.

\section{Method and results}
In this section we carry out a saddle point analysis of the action. 
The resulting field configurations, usually referred to as instantons, provide the leading order contributions to the partition function in the semiclassical limit. 

Depending on the ratio $\mu/\lambda$ there are two qualitatively different configurations, depicted in Fig.\ref{fig3}, of the potential $V(\phi)$: Case A, characterised by two local minima in the interval $[0,4\pi)$, and Case B, characterized by only one global minimum. 
The explicit solutions of the equation, $\delta S'_0 =0$, found in Ref. \cite{49}, greatly simplifies the theoretical analysis. 

\begin{figure}
\includegraphics[width=0.99\columnwidth,clip,angle=0]{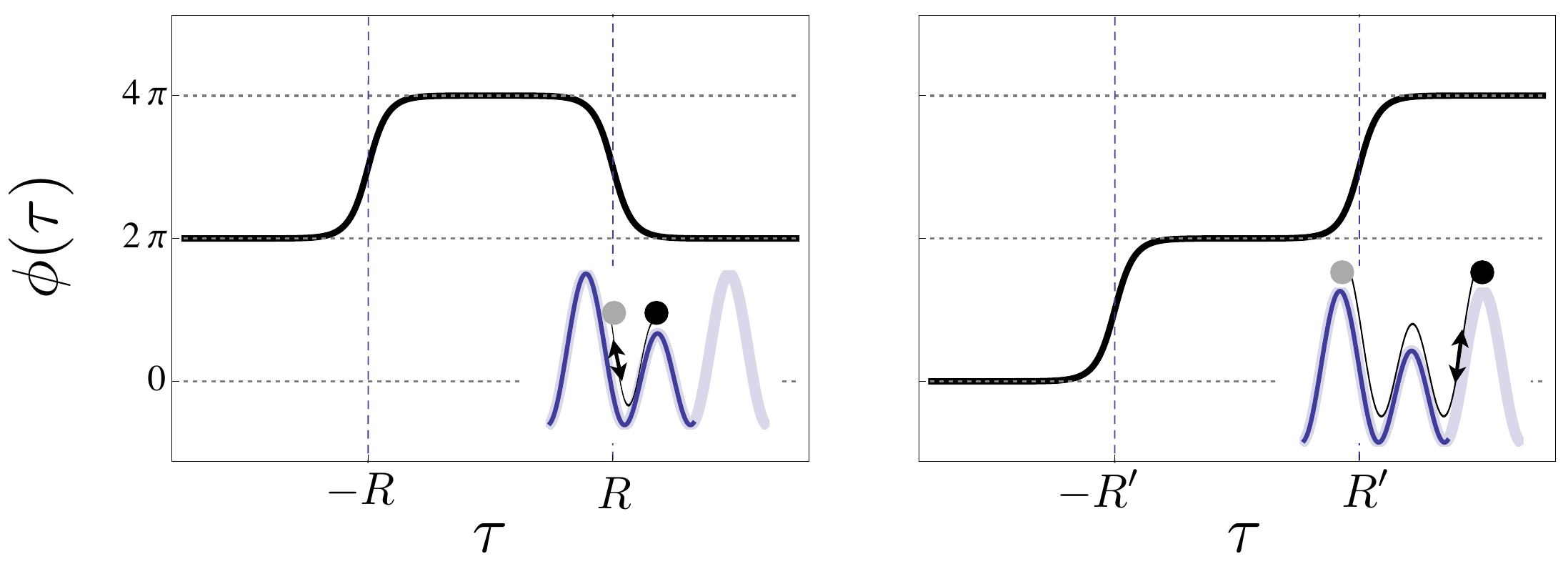}
\caption{(Left) The bounce trajectory for case A. The solution is effectively the sum of an instanton and anti-instanton of the sine-Gordon model. (Right) The trajectory of a single instanton for case A \cite{49}.
} 
\label{fig4}
\end{figure}

Following \cite{49} let us first discuss the bounce-like solution, existing only in case A, 
that starts and finishes at $\phi =2 \pi$. 
For the Wick rotated potential, shown in the left panel inset of Fig. \ref{fig4}, the bounce trajectory corresponds to the phase effectively rolling down the hill and bouncing back at a position where the potential equals that of the local minimum ($\phi= 2\pi$). 
This trajectory is given by, 
\begin{eqnarray}
\phi_{dsG}  &=& \phi_{sG}(\tau + R) + \phi_{sG}(-(\tau-R))
\end{eqnarray}
where $\phi_{sG}(\tau) = 4 \mathrm{tan}^{-1}[e^{m\tau }]$ is the instanton solution of the sine-Gordon model (i.e. the solution of the equations of motion with $ \lambda=0$), 
$R = \frac{1}{m}\mathrm{sinh}^{-1} \left[\sqrt{\frac{4\mu}{\lambda}-1}\right]$ and $m^2 = \mu -\lambda/4$.
These solutions are topologically trivial as they do not cause any phase-slip, namely, the winding number of the phase after one bounce is still zero. 
In the context of Quantum Chromodynamics it has been shown that these bounces contribute to tunnelling but only perturbatively so it is safe to neglect them with respect to the leading non-pertubative contribution to the action \cite{13}. Moreover, ignoring these bouncing trajectories, enables a joint analysis of cases A and B.

Mussardo et al. \cite{49} have also derived the classical instanton solution connecting the minima at $\phi =0$ and $\phi=4\pi$. This solution, shown schematically in the right panel of Fig.\ref{fig4}, is written as a superposition of sine-Gordon instantons:
\begin{eqnarray}
\phi'_{dsG}(\tau) &=& \phi_{sG}(\tau + R')+\phi_{sG}(\tau - R')
\label{dsg}
\end{eqnarray}
with  $ R' =\frac{1}{m'}\mathrm{acosh} \sqrt{\frac{4\mu}{\lambda}+1} $ and $m'^2 = \mu +\lambda/4$. 
%%%%%%%%%%%%%%%%%%%%%%%%%%%%%%%%%%%%%%
%%%%%%%%%%%%%%%%%%%%%%%%%%%%%%%%%%%%%%
As is shown in Fig.\ref{fig4}, the phase spends a time $2R'$ at the local minimum/maximum of the potential ($\phi =2 \pi$) before transitioning to the 
global minimum ($\phi = 4\pi$). 
The expression (\ref{dsg}) also gives the sine-Gordon instantons back in the limit of $\lambda \to 0$ for which   
$R' \to \infty$. This corresponds to the loss of the correlation between the two 
instantons in the double sine-Gordon solution. In this limit therefore the two sine-Gordon instantons 
can be regarded as free \cite{49}.

The results from Pekker et al. \cite{3} that $2\pi$ phase slips are suppressed can clearly 
be seen from the expression for $\phi'_{dsG}$ above, since $\phi'_{dsG}(-\infty)=0$ and 
$\phi'_{dsG}(\infty) = 4\pi$.

Following the treatment of Schmid \cite{22}, we now postulate the following approximate solution, valid in the dilute limit corresponding to large separations between instantons,
\begin{eqnarray}
\Psi_{cl} &=& \sum_{j=1}^ne_j\phi'_{dsG}(\tau-\tau_j)
\end{eqnarray}
where $e_j = +1(-1)$ for instantons (anti-instantons), $n$ is the number of instantons/anti-instantons and $\tau_i$ is the instanton's {\it center of mass}. The condition $\sum_{j=1}^n e_j =0$ ensures that the action is finite. The proposed configuration corresponds to the leading order contribution to the path integral in the limit in which phase slips are still rare events and therefore a linear superposition of well separated instantons is a good approximation to the full path integral. 

To begin the analysis of the instanton contribution to the action, we observe that within the dissipationless action there is no interaction between instantons since we have assumed the typical distance $|\tau_i - \tau_j|$ (with $i,j = 1,...,n $) to be large. In this regime the multi-instanton action can also be approximately given by the factorized expression $S'_{0(n)} \approx n S'_{0(1)}$ with $S'_{0(1)}$ the action of a single instanton. 
We now insert the solution above in the dissipative term of the action and make a further simplifying assumption, valid for large values of $m/\tilde\eta$: the instanton profile is replaced by an Heaviside  $\theta$ function. 

After substituting this ansatz solution in the dissipative term of the action and integrating by parts twice, 
\begin{eqnarray}
S'_{diss} &\approx& 2\tilde\eta(2\pi)^2\sum_{i,j}^ne_ie_j[2\log(\tau_i-\tau_j)+ \nonumber \\ 
& & \log(\tau_i-\tau_j -2R')+\log(\tau_i-\tau_j+2R')].
\end{eqnarray}
This expression can be further simplified assuming $|\tau_i -\tau_j| \gg R'$. 
Neglecting second order terms in $R'$ yelds,
\begin{eqnarray}
S'_{diss} & \approx  & 8 \tilde\eta (2\pi)^2\sum_{i,j}^ne_ie_j\log(\tau_i-\tau_j).
\end{eqnarray}
This result is identical to that obtained for a non-topological Josephson junction with Ohmic dissipation \cite{4, 22} except for the overall rescaling of the pre-factor in the $S'_{diss}$ term. 
The theoretical analysis of Refs.\cite{4,22}, under the assumptions above, yield in our case a critical dissipation $\tilde \eta_c = \frac{E_C}{4\pi^2}$. 
When dissipation is induced by quasiparticle tunnelling the expression of $\tilde\eta_c$ above translates to,
\begin{eqnarray}
R_c = \frac{h}{e^2}
\end{eqnarray}
where $R_c$ is the critical normal state resistance $R_N$.
As a matter of comparison, the critical normal state resistance for the non-topological Josephson junction, $R_c = h/4e^2$, is four times less than in the topological case. 

We note that this result assumes that a small instanton fugacity $z=e^{ - S'_{0(1)}/(8 E_C)} \ll 1$, with,
\begin{eqnarray}
S'_{0(1)} &=& \int  \left\{ \frac{\left[\partial_\tau {\phi'_{dsG}(\tau)}\right]^2}{2} - V[\phi'_{dsG}(\tau)] \right\} d\tau \nonumber,
\end{eqnarray} has a negligible effect on $\tilde\eta_c$. Corrections to $\tilde\eta_c = \frac{E_C}{4\pi^2}$ due to a small $z$ can still be computed systematically within the renormalization group framework of Refs.\cite{4,22}. This correction, as for non-topological JJ, slightly increases $\tilde\eta_c$ though its effect is relatively small in the dilute limit in which the instanton approach is applicable.

%The physical meaning of this factor four difference is directly related to the fact that in a normal superconductor the current is always carried by pairs of electrons of opposite spin while in the topological case it is carried by single low energy fermions. \nota{<- I'm really not sure of this....} Since the charge enters squared in the expression for the resistance a factor four difference is obtained by replacing $2e$ (conventional) by ($e$) topological.
In summary, topological JJ are more robust to phase-slips than the non-topological counterpart. A substantially smaller dissipation is sufficient to stabilize superconductivity in the topological case.

\section{Conclusion}
We have studied the role of dissipation in a topological superconducting junction. 
In general such junction is more robust against fluctuations than the non-topological counterpart. 
As dissipation increases the phase transition to a superconducting state occurs at a critical value of the dissipation which is four times smaller than that expected for a conventional Josephson junction. 
A tentative explanation for this difference is that the current is carried by single fermions (charge $e$) instead of Cooper pairs (charge $2e$) as in conventional JJs. This difference could be used as a robust experimental signature of topological superconductivity. 
These results provide evidence that topological superconductors might be of interest in both quantum information, as a coherent qubit, and also in typical applications of JJ's in situations in which there is no phase coherence in the non-topological JJ because the proliferation of phase slips. 

\acknowledgments AMG was supported by EPSRC, grant No. EP/I004637/1, FCT, grant PTDC/FIS/111348/2009 and
a Marie Curie International Reintegration Grant
PIRG07-GA-2010-268172.

% References %%%%%%%%%%%%%%%%%%%%%%%%%%%%%%%%%%%%%%%%%%%%%%%%%%%%%%%%%%%%%%%%%%%%%

\end{document}